\documentstyle[11pt,newpasp,twoside,graphicx]{article}
\def\edcomment#1{\iffalse\marginpar{\raggedright\sl#1\/}\else\relax\fi}
\reversemarginpar

\begin{document}
\title{The visible matter -- dark matter coupling}

 \author{Renzo Sancisi}
\affil{
Osservatorio Astronomico, Via Ranzani 1, I-40127 Bologna, Italy\\
Kapteyn Astronomical Institute, PO Box 800, NL-9700 AV Groningen, The Netherlands
}

\begin{abstract}
In the inner parts of spiral galaxies, of high or low surface brightness, 
there is a close correlation between rotation curve shape and light 
distribution. For any feature in the luminosity profile there is a corresponding 
feature in the rotation curve and vice versa. 
This implies that the gravitational potential is strongly correlated with 
the distribution of luminosity: either the luminous mass dominates or there is a 
close coupling between luminous and dark matter.
In a similar way, the declining rotation curves observed in the outer parts of high 
luminosity systems are a clear signature of the stellar disk which either dominates 
or traces the distribution of mass. 

The notion that the baryons are dynamically important in the centres of galaxies,
including LSBs, undermines the whole controversy over the cusps in CDM halos and the 
comparison with the observations. If the baryons dominate in the central regions of all
spirals, including LSBs, how can the CDM profiles be compared with the observations? 
Alternatively, if the baryons do not dominate but simply trace the DM distribution,
why, in systems of comparable luminosity, are some DM halos cuspy 
(like the light) and others (also like the light) are not?
\end{abstract}

\section{Introduction}
The relation between the distribution of light and the shape of rotation curves
has been at the centre of the discussion on the interpretation of rotation curves and 
on the presence and the amount of dark matter in spiral galaxies.
Burstein and Rubin (1985) discussed the systematic properties of rotation curves and 
concluded that for a given morphological type the shapes of rotation curves vary 
systematically with luminosity; these shapes are, however, similar for different 
morphologies and 
therefore the form of the gravitational potential is not correlated with the form 
of the light distribution or the morphological type. This was used as an argument in 
favour of the presence of dark matter also in the inner regions of galaxies.

In contrast to this conclusion, the success of the 
maximum disk hypothesis in the analysis of rotation curves (Kalnajs 1983, Van Albada 
and Sancisi 1986, Kent 1986, 1987, Broeils 1992, Palunas and Williams 2000) implies that 
within the optical disk either the mass of dark matter is small or that its 
distribution is very closely coupled to the distribution of luminous matter. 
The maximum disk analysis was carried out on observations of predominantly high 
surface brightness (HSB)
systems. Recent work on low luminosity and low surface brightness (LSB) galaxies has shown 
that for the inner parts it is possible to obtain maximum disk fits as good as for 
the more luminous 
systems (Verheijen 1997, Swaters 1999). The main problem with the maximum disk in these 
systems is caused by the required large M/L ratios which would seem to argue in favour 
of DM being dominant everywhere.
	
Persic and Salucci (1991) and Persic, Salucci and Stel (1996) have discussed the relation 
between luminosity and rotation curves and have emphasized the strong dependence 
on luminosity for both the shape and the amplitude of the rotation curve. They have 
come to the conclusion, based on a large galaxy sample, that the observed curves have 
a universal shape, dependent only on total luminosity. However, Verheijen (1997) finds 	
that roughly one third of the observed HI rotation curves from his Ursa Major 
sample of galaxies deviate noticeably from the "universal rotation curve" shape.	
	
	A new interest in the question of the shape of rotation curves and of the mass 
distribution in the central parts of spirals has recently been originated by the debate 
on the presence of cusps as predicted by the Cold Dark Matter simulations on structure 
formation (Navarro, Frenk and White 1997). Naturally the low luminosity and LSB 
galaxies, believed to be DM dominated, are taken as the best and most suitable for the 
test and their central rotation curves are thoroughly investigated. It is clear that 
if baryons dominate the mass distribution in the central regions of galaxies this would 
affect the whole discussion on CDM cusps and the comparison with the rotation curves;
however, this could also reveal something about the distribution of DM. 
The problem is then to properly understand how cores are formed. 
If, on the other hand, DM dominates everywhere and closely follows the distribution of 
the light, then the question is why some dark halos are centrally peaked like 
the stars while others are not (again mimicking the light).

\section{The low luminosity and low surface brightness systems}

Are the low luminosity and LSB galaxies really dark matter 
dominated everywhere, also within their optical radii? This is not necessarily the case. 
The rotation curve of the stellar disk calculated from the fotometry can be scaled up 
to match the observed curve in the inner parts, usually within a few kiloparsec, and 
the quality of this maximum disk fit is not inferior to that obtained for 
the HSB galaxies (Swaters, 1999).  
However, the required value for the mass/luminosity ratio is 
high, up to about 15 in the R-band, and this is the main reason for believing that the 
low luminosity and LSB galaxies are DM dominated. It is remarkable, however, that in 
the inner parts the curve from the maximum disk solution matches the observed curve. 

This suggestion of a tight link between DM and stellar component in the central 
parts of these low luminosity 
and LSB systems is strongly reinforced by the following observation: every time there is 
a feature in the radial light distribution the rotation curve shows a corresponding 
feature. For instance, to a central concentration of light corresponds an excess of 
rotational velocity. Conversely, to a feature (steep gradient, bump) in the rotation 
curve always corresponds an excess in the luminosity profile. This has been often 
pointed out for the HSBs (e.g. Van Albada and Sancisi 1986, Palunas and Williams 2000),
but it is clear now that it also holds for the LSBs. 
To my knowledge there is no exception to this rule.
Fig. 1 shows the luminosity profile (Swaters 1999) and the rotation curve 
(Zwaan, Van der Hulst and Bosma 2003) of the LSB galaxy NGC 3657 
(UGC 6406). This galaxy is dominated by a very bright central concentration. 
The rotation curve shows a steep rise and a peak inside 50 arcsec (5 kpc) 
instead of the slow rise characteristic of LSB systems. This feature in the rotation 
curve corresponds to the strong central concentration in the luminosity profile.
Maximum disk fits give M/L values (R-band) between 2.9 and 4.2  for the central
concentration and between 13.8 and 14.7 for the outer disk (Zwaan et al. 2003).
A very similar case is that of NGC 5963 (UGC 9906) as shown by the comparison of the 
photometry (Swaters 1999) and the rotation curve (Zwaan et al. 2003). 
At still lower luminosities, a good example is that of NGC 5585 where the luminosity 
profile shows the presence of a small central concentration of light and the optical 
rotation curve (Blais-Ouellette et al. 1999) reveals a corresponding bump.  

In the past there have been cases of measured rotation curves differing considerably 
from those predicted by the light profile, but the discrepancies have turned out to 
be apparent only,  
and have been traced to a problem with the light profile or with the rotation curve. 
This is well illustrated by the case of the late-type, edge-on spiral NGC 5907. Its 
observed rotation curve rises rapidly near the centre, whereas the rotation curve 
predicted from the light profile (Van der Kruit and Searle 1981) rises much less 
steeply (see figure 7 in Sancisi and Van Albada 1987). This seemed  
to indicate a clear need for an additional component, in the central parts, of matter 
unrelated to the light. Van Albada and Sancisi (1986) suggested that the discrepancy 
might be caused by a problem in the light profile and that the "missing material" might 
well be the luminous material obscured by the dust. Several years later, this was proved  
to be right by Barnaby and Thronson (1994) who used  H-band surface photometry and 
successfully modelled the observed rotation curve in the inner parts of NGC 5907.

The relation between the rotation curve shape and the central concentration of light 
has been investigated quantitatively by Swaters (1999) and more recently  
by Swaters and Sancisi (2003) for a large 
sample of spiral galaxies including late-type dwarfs and LSBs. A clear relation is found
between the central concentration of light measured from the luminosity profiles and the 
steepness in the rise of the rotation curve in the inner parts.   
Similarly, Verheijen (1997) has found a correlation between the shapes of the rotation 
curves and the compactness of the luminous disk: for galaxies of the same luminosity a 
more compact distribution corresponds to a steeper rise in the rotation curve. This 
is in contrast with Burstein and Rubin's (1985) conclusions mentioned above.
At any rate, in order to confirm and to investigate further the tight link between light 
distribution and rotation it is important to observe a larger sample of low luminosity and 
LSB systems with a central light concentration.

\begin{figure}
\centerline{
\includegraphics[height=0.40\textheight]{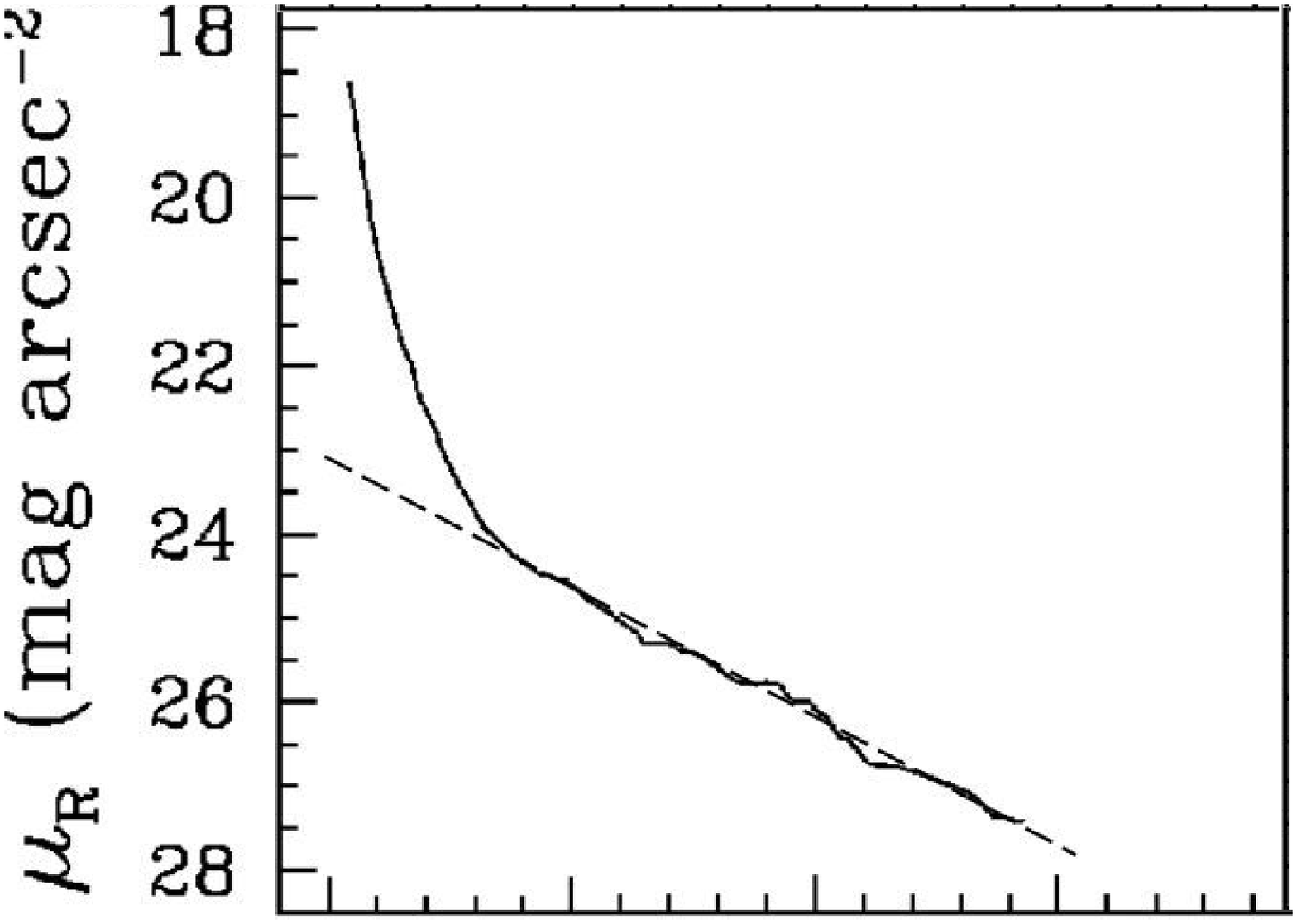}}
\centerline{
\includegraphics[height=0.47\textheight]{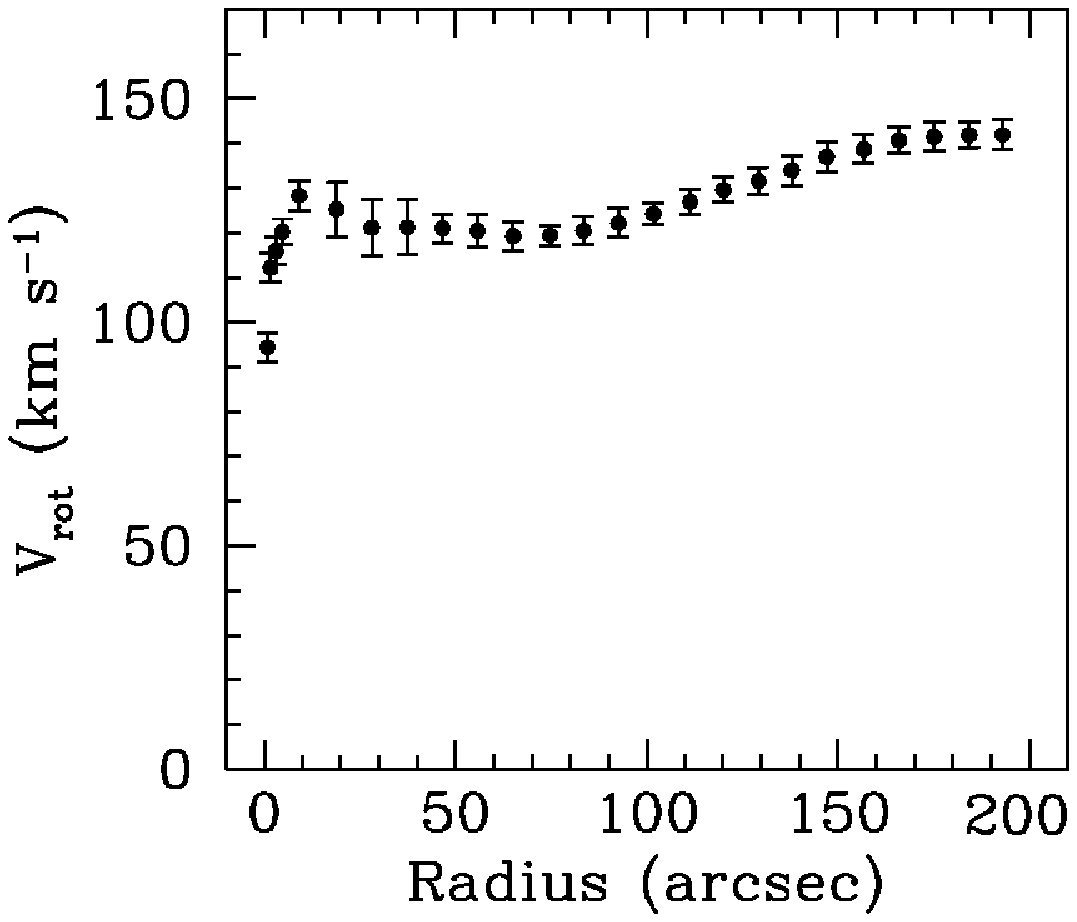}}
\caption{The low surface brightness galaxy NGC 3657 (UGC 6406). Top: Radial surface 
brightness profile in R-band (Swaters, 1999). The dashed line shows an exponential 
fit to the outer disk. Bottom: HI rotation curve (Zwaan et al., 2003). 10 arcsec=1 kpc.}
\end{figure}

\section{The high luminosity systems}

In high surface brightness systems the maximum disk model adopted for the 
decomposition of rotation curves provides good fits to the rotation curves in the 
bright inner parts of the disks (see references in section 1). The very luminous 
galaxies provide a very 
interesting case for the comparison of the distribution of light and DM. 
This is illustrated in the paper by  Noordermeer et al. (this volume). 
Casertano and Van Gorkom (1991), Persic and Salucci (1991) and Broeils (1992) 
pointed out that galaxies with high luminosity (and sometimes very compact) 
disks have declining rotation curves in their outer parts. 
Fig. 2 shows the results for the luminous spiral galaxy NGC 5055 recently 
obtained by Battaglia et al. (2003). The plot shows the observed HI rotation curve 
and the standard decomposition with maximum disk model (derived from the F-band photometric 
profile shown in the top panel) and isothermal halo. The rotation curve has the steep 
rise characteristic of very luminous HSBs and shows a bump in the inner regions, 
around 2 kpc. The bump is present on both galaxy sides (receding and approaching) 
and completely symmetrical. The photometric profile shows a bump at the same 
radius and indeed the maximum disk rotation curve calculated from it reproduces the 
bump in the observed rotation curve perfectly. This remarkable correspondence between 
the distribution of light and the rotation is very representative for what
is found in many galaxies.
 
In the outer parts, around the Holmberg radius R(Ho), 
the rotation curve of NGC 5055 declines by 
about 25 km/s and remains flat out to the last measured point at 40 kpc.
The decomposition of the rotation curve shows that the maximum disk model 
(with isothermal or NFW profile) reproduces the shape of the rotation curve also 
in the outer parts. Solutions with NFW profile 
and a disk below maximum are not satisfactory.  With a isothermal halo, instead, it is 
possible to obtain acceptable fits also with sub-maximal disks and to  
set a firm lower limit to the mass/luminosity ratio of the stellar disk. 
The derived "minimum" disk is rather massive: it contributes about 63 percent of  
the maximum rotational velocity and has M/L=0.8 (F-band). 
Similar results have been 
obtained by Bottema and Verheijen (2002) for NGC 3992 and by Noordermeer et al. 
(paper in this volume) for a number of luminous early-type galaxies.   

It is interesting to note that these "minimum" disks are close to
the "Bottema" or similar sub-maximal disks which are obtained   
from the measurements of stellar velocity dispersion (Bottema 1993, Kregel 2003).
These contribute on average about 
60 percent of the observed maximum rotational velocity (as compared  
to the about 90 percent of the maximum disks). Clearly, although not maximal, 
they still are rather massive.

\begin{figure}
\centerline{
\includegraphics[height=0.38\textheight]{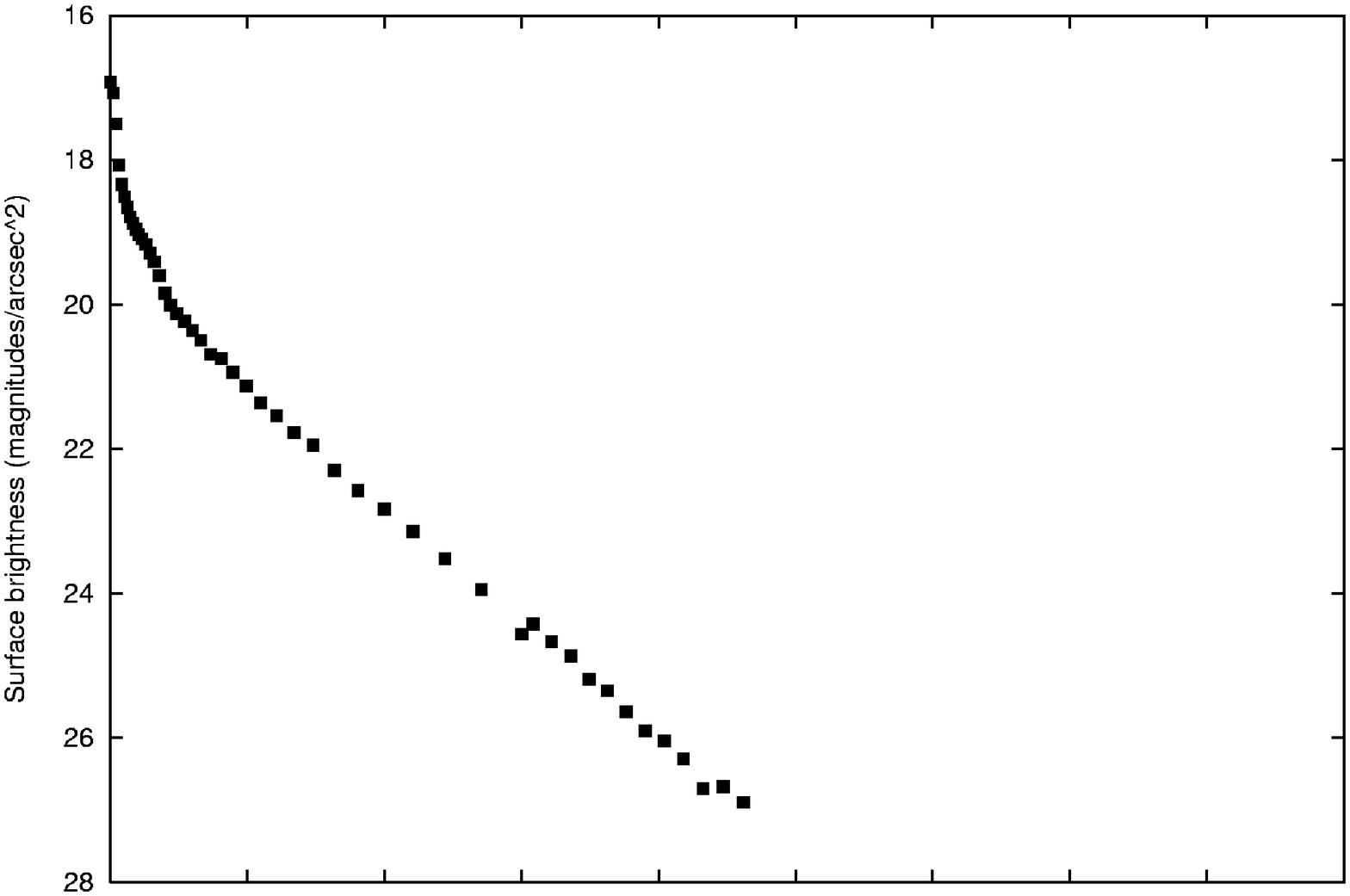}}
\vglue 0.1cm
\centerline{
\includegraphics[height=0.40\textheight]{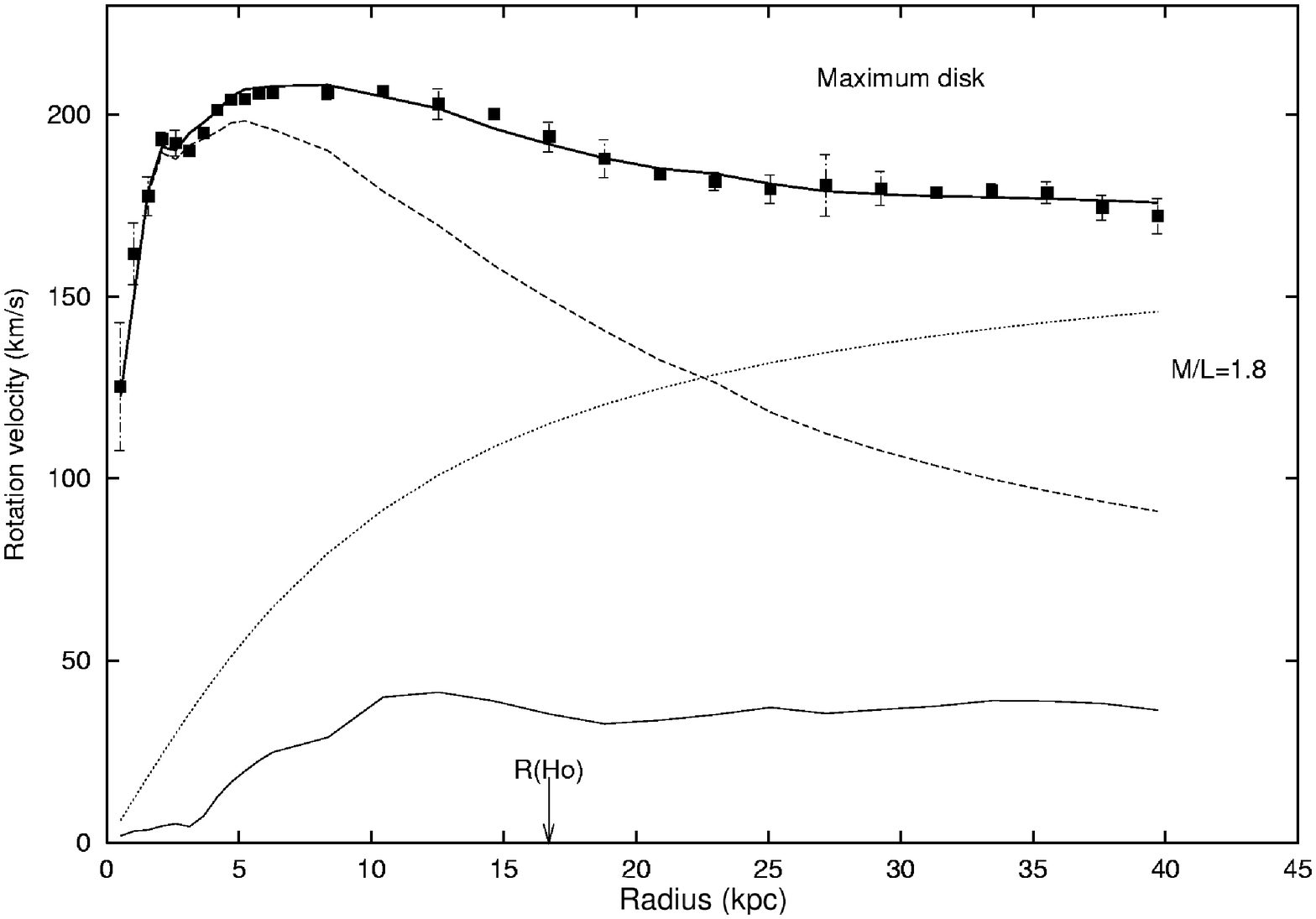}}
\caption{The high luminosity spiral NGC 5055. Top: Radial surface brightness profile in
F-band. Bottom: Mass model with maximum disk solution (constant M/L=1.8). 
The observed HI rotation curve is
shown by the filled squares with error bars, the maximum stellar disk by the dashed line,
the isothermal DM halo by the dotted line and the HI disk by the full thin line. The full 
thick line shows the quadratic sum of the contributions from all components}
\end{figure}

\section{Main points}

\hskip\parindent 1. There is a striking correspondence between the shape of the rotation 
curves and the shape of the radial distribution of luminosity in spiral galaxies.
In the central parts, for objects of the same luminosity, a more compact
distribution of light gives rise to a steeper rotation curve.
Over the inner bright part of the optical disk the maximum disk hypothesis provides 
satisfactory fits to the rotation curves. In other words, 
the correspondence between rotation curve and distribution of light is close and detailed. 
There is a simple rule -to an excess of light 
corresponds an excess of rotation and vice versa- which seems to apply to all spiral 
galaxies including the late-type dwarfs and the low surface brightness objects. 
I do not know of any counterexample. Clearly, this needs to be verified with measurements 
of rotation curves, in particular of dwarf galaxies and LSBs with central light
concentrations. 

2. A special case of such remarkable correspondence between light distribution and shape 
of the rotation curve is that of the high luminosity, early-type galaxies.
In the outer parts of these systems, beyond their bright optical disk, 
the rotation curve declines somewhat and remains flat further out to the last measured point.
This behaviour strongly suggests that the disk potential dominates in the inner parts and the
DM halo in the outer parts. These cases are important because they permit to set firm 
lower limits to the disk mass. The values derived for the M/L ratios indicate that the 
'minimum" disk is rather heavy and the stellar component is indeed important over the 
optically bright part of the galaxy.

3. The rule suggested here can be seen perhaps as a Tully-Fisher kind of relation, between
the distributions of light and the run of rotational velocities. Whilst the T-F is a 
relation between global properties  -total luminosity and amplitude of the 
rotation curve-, this is a point-to-point relation between luminosity and rotational 
velocity over the galaxy.

4. The unavoidable conclusion from the observed correspondence between the shapes of the 
rotation curves and those of the luminosity profiles is that the gravitational 
potential is strongly correlated with the distribution of luminous matter: either the 
luminous mass dominates or there is a close coupling between luminous and dark matter.

5. Clearly these results bear on the debate on cusps in the mass profiles of the central 
regions of disk galaxies as predicted by CDM simulations. The amazing fact is that when 
the rotation curve indicates a concentration of mass -a cusp-, such a cusp shows up in 
the light. Then the following questions arise: if the baryons indeed dominate 
in the central regions of all spirals, LSBs included, how can the CDM profiles be compared 
with the observations? 
If, on the other hand, the baryons do not dominate but trace the DM distribution, why, in
systems of comparable luminosity, are some DM halos cuspy (following the visible matter) 
and others (also following the visible matter) are not?

\acknowledgments

I thank Filippo Fraternali, Luca Ciotti and Rob Swaters for stimulating discussions and 
helpful comments on the manuscript and Giuseppina Battaglia for providing 
the figure for NGC 5055.

\end{document}